\documentclass[11pt]{article} 
\pdfoutput=1 

\usepackage{jheppub} 

\usepackage[utf8]{inputenc}
\usepackage[english]{babel}
\usepackage{amsmath,amssymb,amsbsy,amstext,amsthm,booktabs,simplewick,exscale,relsize,slashed,graphicx,amsfonts,upgreek, xcolor,subcaption}
\usepackage{multirow,color,dcolumn,bm,enumerate}
\usepackage{feynmp-auto,axodraw2,tikz}
\usepackage{ulem}
\usepackage{comment}
\usepackage{cancel}
\usepackage[compat=1.1.0]{tikz-feynman}
\usepackage{standalone}
\usepackage{array}
\usepackage[most]{tcolorbox}
\usepackage{tabularray}
\UseTblrLibrary{booktabs}

\usepackage{mathtools}

\def\figureautorefname~#1\null{Fig.\,#1\null}

\def\equationautorefname~#1\null{Eq.\,(#1)\null}

\usepackage[dvipsnames]{xcolor}

\usepackage{hyperref}

\usepackage[capitalise,noabbrev,nameinlink]{cleveref}
\title{Positivity Bounds in $\mathcal N$=1 Supersymmetry}
\author[a]{Jingxuan Bu,}
\author[a,b]{Jiayin Gu,}
\author[a]{Runqing Wang}

\affiliation[a]{Department of Physics and Center for Field Theory and Particle Physics,\\ Fudan University, Shanghai 200438, China}
\affiliation[b]{Key Laboratory of Nuclear Physics and Ion-beam Application (MOE), \\ Fudan University, 
Shanghai 200433, China}
\emailAdd{jxbo25@m.fudan.edu.cn}
\emailAdd{jiayin\underline{~}gu@fudan.edu.cn}
\emailAdd{runqing\underline{~}wang@fudan.edu.cn}

\abstract{For a low-energy Effective Field Theory (EFT) to admit a consistent ultraviolet (UV) completion, it must adhere to the fundamental principles of locality, unitarity, analyticity, and Lorentz invariance. This leads to positivity constraints on certain Wilson coefficients via dispersion relations of $2\rightarrow2$ forward elastic amplitudes, which carry significant implications for both theoretical consistency and experimental phenomenology. In this work, we extend this bootstrap framework to $\mathcal{N}=1$ supersymmetric theories, where the super-Poincar\'e algebra relates the bosonic and fermionic degrees of freedom within a single supermultiplet. Supersymmetric Ward identities (SWIs) enforce exact linear relations among component scattering amplitudes. Consequently, the Wilson coefficients associated with different $2\rightarrow2$ processes are mutually constrained, rendering their respective positivity bounds dependent. Focusing on dimension-8 operators, we explicitly show the positivity bounds on quartic interactions involving scalars, fermions and gauge fields, and demonstrate how they are related as a direct consequence of the SWIs. Furthermore, we expect that the algebraic nature of this framework naturally extends to loop-level superamplitudes and higher-point processes, providing a unified bootstrap perspective on supersymmetric EFTs.}

\begin{document}
\maketitle
\setcounter{page}{2}

\section{Introduction}
Effective field theory (EFT) provides a systematic framework for encoding unknown high-energy physics into low-energy observables. From a bottom-up perspective, higher-dimensional operators parameterize our ignorance of physics at a UV or cutoff scale $\Lambda$. Adopting the Lagrangian approach, a typical EFT is organized as an  expansion in terms of operator dimensions:
\begin{equation}
    \mathcal{L}_{\rm EFT}=\mathcal{L}_4+\sum_{n\geq5}\sum_{i}\frac{c_{ni}\mathcal{O}_{ni}}{\Lambda^{n-4}}\,,
\end{equation}
where $\mathcal{L}_4$ denotes the renormalizable part of the theory; $\mathcal{O}_{ni}$ represents an operator of mass dimension $n$, the index $i$ runs over all independent operators of that dimension, and $c_{ni}$ are the associated dimensionless Wilson coefficients. While these Wilson coefficients are generally treated as free parameters to be determined experimentally, they are not entirely unconstrained. Since an EFT emerges as the low-energy limit of a consistent ultraviolet (UV) theory, it must respect the fundamental axioms of Quantum Field Theory (QFT): locality, unitarity, analyticity, and Lorentz invariance. These principles yield rigorous theoretical constraints such as positivity bounds, which significantly restrict the viable parameter space. This intersection of theoretical consistency and phenomenological testability has motivated extensive research \cite{Distler:2006if,Manohar:2008tc,Nicolis:2009qm,Bellazzini:2014waa,Bellazzini:2015cra,Bellazzini:2016xrt,deRham:2017avq,deRham:2017zjm,Bellazzini:2017bkb,Bellazzini:2017fep,deRham:2017xox,deRham:2018qqo,Bellazzini:2019xts,Bellazzini:2019bzh,Wang:2020jxr,Zhang:2020jyn,Alberte:2020jsk,Tokuda:2020mlf,Bellazzini:2020cot,Tolley:2020gtv,Trott:2020ebl,Herrero-Valea:2020wxz,Sinha:2020win,Caron-Huot:2020cmc,Alberte:2020bdz,Arkani-Hamed:2020blm,Li:2021lpe,Caron-Huot:2021rmr,Bern:2021ppb,Aoki:2021ckh,Chiang:2021ziz,Guerrieri:2021tak,Henriksson:2021ymi,Davighi:2021osh,Arkani-Hamed:2021ajd,Du:2021byy,Alberte:2021dnj,Bellazzini:2021oaj,Caron-Huot:2022ugt,Chiang:2022jep,deRham:2022hpx,Chiang:2022ltp,Herrero-Valea:2022lfd,CarrilloGonzalez:2022fwg,Haring:2022sdp,Fernandez:2022kzi,Riembau:2022yse,Hamada:2023cyt,Hong:2023zgm,Bellazzini:2023nqj,CarrilloGonzalez:2023cbf,Xu:2023lpq,Bertucci:2024qzt,Bittar:2024xuc,Caron-Huot:2024tsk,Beadle:2024hqg,Ye:2024rzr,Buoninfante:2024ibt,Xu:2024iao,Wan:2024eto,Peng:2025klv,Chang:2025cxc,Beadle:2025cdx,Tokareva:2025rta,Desai:2025alt,Liao:2025npz,deRham:2025vaq,Low:2009di,Bellazzini:2018paj,Zhang:2018shp,Bi:2019phv, Remmen:2019cyz,Remmen:2020vts,Fuks:2020ujk,Yamashita:2020gtt,Gu:2020ldn,Bonnefoy:2020yee,Remmen:2020uze,Gu:2020thj,Chala:2021wpj, Zhang:2021eeo,Azatov:2021ygj,Li:2022tcz, Li:2022rag, Li:2022aby,Ghosh:2022qqq,Remmen:2022orj,Chen:2023bhu, Gu:2023emi, Davighi:2023acq,  Chala:2023jyx, Chala:2023xjy, Altmannshofer:2023bfk, DasBakshi:2023htx,Ye:2025zhs,Bellazzini:2025bay}. 

The most straightforward positivity bounds follow from the dispersion relation of a $2\rightarrow 2$ forward elastic amplitude at the tree level. In the massless limit $(m\to0)$, the operator–amplitude correspondence \cite{Shadmi:2018xan,Ma:2019gtx,Aoude:2019tzn,Durieux:2019eor,Durieux:2019siw,Gu:2020thj} identifies Wilson coefficients with the coefficients of the local contact terms in the low-energy expansion of the scattering amplitude. A cornerstone result is that the second derivative of the $2\rightarrow 2$ elastic amplitude $\mathcal{A}_{2\rightarrow2}(s,t)$\footnote{In a general $2\rightarrow 2$ scattering process, the Mandelstam variables are related via $s+t+u=\sum_{i=1}^4 m_i^2$.} with respect to the Mandelstam variable $s$ in the forward limit $t\rightarrow 0$ must be non-negative. This condition follows directly from the optical theorem and the positivity of total cross sections. Research has predominantly focused on dimension-8 operators, as they offer both theoretical tractability and phenomenological accessibility: the resulting bounds are generally testable at colliders and serve as a critical consistency check for any viable EFT.

Positivity constraints persist beyond tree level. As demonstrated in Refs.~\cite{Bellazzini:2020cot,Herrero-Valea:2020wxz,Arkani-Hamed:2020blm, Arkani-Hamed:2021ajd,Chala:2021wpj,Bellazzini:2021oaj,Li:2022aby, Chala:2023jyx,Caron-Huot:2024tsk,Beadle:2024hqg,Ye:2024rzr,Peng:2025klv,Chang:2025cxc,Beadle:2025cdx,Tokareva:2025rta,Desai:2025alt,Liao:2025npz,Ye:2025zhs}, loop corrections must be carefully incorporated for the bounds to hold rigorously. A consistent derivation at the loop level requires accounting for lower-dimensional operators (dimensions 4 and 6) \cite{Ye:2024rzr,Ye:2025zhs} to avoid apparent violations of tree‑level constraints. Furthermore, when multiple Wilson coefficients are constrained simultaneously, the allowed region forms a convex geometry known as the ``positivity cone'' or the ``EFT-Hedron'' \cite{Li:2021lpe,Li:2022tcz,Chen:2023bhu,Peng:2025klv,Chiang:2022ltp,Arkani-Hamed:2020blm}. Recent work \cite{Li:2021lpe,Du:2021byy,Elvang:2026,Jeong:2026xzk} has extended these ideas to broader interaction classes, including higher-point (more than four external legs) and higher-dimensional (above mass dimension eight), where gluing and factorization techniques motivate the notion of ``multipositivity bounds''.

Supersymmetry, as a spacetime symmetry, relates bosonic and fermionic degrees of freedom by enlarging the Poincar\'e group to the super-Poincar\'e group. In the Wess-Zumino model, for instance, the Yukawa coupling $y$ is related to the coupling constant of the quartic $|\phi|^4$ interaction $\lambda$, {\it i.e.}, $\lambda \propto y^2$. This algebraic relation raises a natural question: how do component positivity bounds correlate when the underlying scattering amplitudes are no longer independent, but are instead related by supersymmetry? While similar analyses have been pursued in $\mathcal N=8$ supergravity~\cite{Bellazzini:2015cra,Bellazzini:2025shd}, this work focuses on the $\mathcal N=1$ case. From an off-shell perspective, these non-trivial relations arise directly from the closure of the supersymmetry algebra, to ensure manifest supersymmetric invariance. The same principle applies to higher-dimensional interactions, where the coefficients of different component interactions are related by imposing supersymmetry, as encoded in the so-called supersymmetric  Ward identities. Since we mainly focus on the EFT construction, an operator basis at dimension-8 turns out to be helpful and was proposed in \cite{Delgado:2022bho,Delgado:2023ogc}, covering both formal constructions and phenomenological topics.

More explicitly, we investigate how supersymmetry governs the correlations among Wilson coefficients, contrasting with the non-supersymmetric case. If positivity independently constrains two couplings, do those bounds remain distinct once supersymmetry relates the corresponding component amplitudes? For example, consider the dimension-8 operators $\mathcal{O}(\partial^4\phi^4)$ and $\mathcal{O}(\partial^3\phi^2\psi^2)$, whose coefficients are bounded by $C_{\partial^4\phi^4}\geq0$ and $C_{\partial^3\phi^2\psi^2}\geq0$. When these operators reside in the same supermultiplet, supersymmetry enforces exact linear relations among their coefficients. Consequently, the positivity condition for one bound automatically implies constraints on its supersymmetric partners, allowing one to infer bounds without independent re-derivation. We invoke supersymmetric Ward identity as a powerful tool for our purposes and demonstrate how it relates different on-shell component amplitudes in later sections.

In this paper, we study amplitudes involving different species (scalars, fermions, and gauge bosons) that are related by supersymmetry in a purely supersymmetric theory. Specifically, we focus on the simplest case $\mathcal{N}=1$, leaving the more constrained $\mathcal{N}=2,4$ theories for future investigation. We demonstrate that the corresponding Wilson coefficients are no longer independent at the component level but instead are subject to unified positivity requirements. By constructing a general four-point superamplitude, we show that component amplitudes are intrinsically correlated. As concrete examples, we also explicitly write down the tree-level positivity bounds for several representative models and check the implications of supersymmetry on these bounds. We emphasize that not all Wilson coefficients are constrained by positivity (in both non-supersymmetric and supersymmetric theories), and our analysis focuses exclusively on those coefficients that are subject to rigorous bounds.

The rest of this paper is organized as follows: In Section \ref{sec: review of positivity bounds}, we briefly review the positivity bounds for the $2\rightarrow2$ scattering process and their implications for the Wilson coefficients. We then generalize to the $\mathcal N=1$ supersymmetric case in Section \ref{sec: supersymmetric positivity}. Specifically, we start with the general four-point superamplitude in Section \ref{sec: supersymmetric 4-pt}; then we take the forward limit and impose crossing symmetry in Section \ref{sec: supersymmetric forward limit} to show the positivity of supersymmetric four-point amplitudes in Section \ref{sec: positivity of  supersymmetry}. After establishing the theoretical framework, we enumerate several $2\rightarrow2$ contact interactions and their supersymmetric partners in Section \ref{sec: examples}, and show that the results are consistent with our earlier derivations. We review the massless spinor-helicity formalism and specify the notation and conventions in Appendix \ref{app:spinor-helicity}. Then we show that applying supersymmetric Ward identities leads to the same result and give an example in Appendix \ref{appendix: supersymmetric Ward Identity}.

\section{A brief review of positivity bounds}\label{sec: review of positivity bounds}
Positivity bounds provide a rigorous connection between low-energy effective field theories (EFTs) and their ultraviolet (UV) completions. These bounds arise from the observation that any consistent UV theory dictates specific positivity conditions for forward elastic $2\to2$ scattering amplitudes $\mathcal{A}_{ab\to ab}(s,t)$ evaluated in the forward limit $t\rightarrow0$. When expanded at low energies, the polynomial dependence of these amplitudes on the Mandelstam variable $s$ is controlled by the Wilson coefficients of higher-dimensional operators, which in turn implies nontrivial positivity constraints on specific combinations of these coefficients. While multiple derivations exist in the literature, we present the standard dispersion-relation approach below.

Consider a generic $2 \to 2$ elastic scattering process $a b \to a b$ with amplitude $A(s,t)$, where $s$ and $t$ are the Mandelstam variables. For simplicity, we assume particles $a$ and $b$ both have mass $m$, while the Mandelstam variable $u$ is fixed by $s+t+u=4m^2$. The forward amplitude is defined by taking the limit $t \to 0$,
\begin{equation}
\mathcal A^{a b \to a b}(s) \equiv \mathcal A(a b \to a b)|_{t\to 0}\,,
\label{eq: A defination}
\end{equation}
which depends solely on $s$. 
\begin{figure}[t]
\centering
\includegraphics[width=0.6\textwidth]{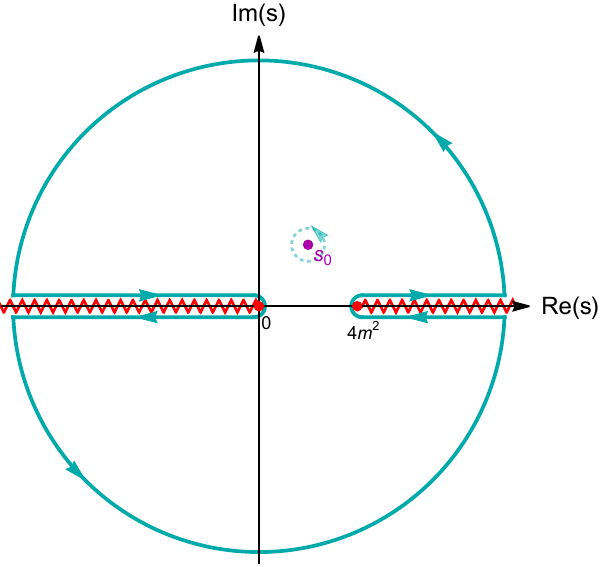}
\caption{Integration contours in the complex $s$-plane at fixed $t=0$. 
The point $s=s_{0}$ is chosen as the reference point for the Cauchy integral.}
\label{fig:contour}
\end{figure}
We define the contour integral of $\mathcal A(s)/(s-s_0)^{n+1}$ along a closed contour $\Gamma$ in the complex $s$-plane as
\begin{equation}
\Sigma^{\mathrm{IR}}\equiv\frac{1}{2\pi i}\oint_{\Gamma}
\frac{ds \mathcal A(s)}{(s-s_0)^{n+1}}\,.
\end{equation}
For simplicity, we assume that there are no isolated poles in
$0<s<4m^2$. The point $s=s_0$ lies in this analytic interval,
and $\Gamma$ encloses the pole at $s=s_0$ introduced by the denominator, as illustrated in Fig.~\ref{fig:contour}.

Using Cauchy's integral theorem, we deform the contour outward to the large circle shown in Fig.~\ref{fig:contour}. Under the Froissart--Martin bound,
\begin{equation}
|\mathcal A(s\to\infty)| < \text{const} \cdot s \log^2 s\,,
\end{equation}
which holds in any local quantum field theory, and the boundary contribution vanishes for $n>1$,
\begin{equation}
\lim_{s\to\infty} \frac{|\mathcal A(s)|}{s^{n+1}} = 0\,.
\end{equation}
Consequently, for $n>1$, the deformed contour reduces to integrals along the branch cuts on the real axis. For equal-mass elastic scattering, the $s$-channel cut starts at the two-particle threshold $s=4m^2$, while the crossed-channel cut extends over $s\leq 0$, since $u=4m^2-s$ in the forward limit. Writing the result in terms of the discontinuity across these cuts, $\mathrm{Disc}\,\mathcal A(s)\equiv\mathcal A(s+i\epsilon)-\mathcal A(s-i\epsilon)$, we obtain
\begin{equation}
\Sigma^{\mathrm{IR}}=\frac{1}{2\pi i}\left(\int_{4m^2}^{\infty} ds+\int_{-\infty}^{0} ds
\right)\frac{\mathrm{Disc} \mathcal A(s)}{(s-s_0)^{n+1}}\,.
\end{equation}

The integral along the $u$-channel branch $s\in(-\infty,0)$ involves unphysical values of $s$. However, it can be rewritten in terms of a physical amplitude involving antiparticles (denoted by a bar, as $\bar{b}$), and related to the original amplitude by crossing symmetry. In particular, crossing particles $2$ and $4$ in the forward elastic limit $t=0$ implies, 
\begin{equation}
A^{ab}(s)=A^{a\bar{b}}(u)\,,\qquad u = -s + 4m^2\,.
\end{equation}
Furthermore, since scattering amplitudes are real-analytic functions of complex variables $\mathcal A(s)^* = \mathcal A(s^*)$, the discontinuity across the branch cut is proportional to the imaginary part. As a result, we obtain the following dispersion relation connecting IR and UV data:
\begin{equation}
\Sigma^{\mathrm{IR}}=\int_{4m^2}^{\infty} \frac{ds}{\pi}
\left[\frac{\mathrm{Im} \mathcal A^{ab}(s)}{(s-s_0)^{n+1}}+
\frac{(-1)^n\mathrm{Im}  \mathcal A^{a\bar{b}}(s)}{(s-4m^2+s_0)^{n+1}}\right]\,.
\end{equation}
Unitarity of the $S$-matrix implies the optical theorem, which relates the forward amplitude's imaginary part to the total cross section:
\begin{equation}
\mathrm{Im} \mathcal A(s)=s \sqrt{1 - \frac{4m^2}{s}}\sigma^{\mathrm{tot}}(s)> 0\,,
\end{equation}
where $\sigma^{\mathrm{tot}}(s)$ is the total cross section. Therefore, for $0 < s_0 < 4m^2$, the integrand is strictly positive for any interacting theory.  
One then arrives at the rigorous positivity bound,
\begin{equation}
\Sigma^{\mathrm{IR}} > 0 \quad\quad \mbox{for even } n \,. \label{eq:pos}
\end{equation}
For the local tree-level four-point interactions considered in this work, the case $n=2$ probes the coefficient of the $s^2$ term in the low-energy expansion of the forward amplitude. Such terms are generated by dimension-8 operators, which will be the focus of the subsequent sections. We also assume that the $m\to 0$ limit can be taken smoothly so that Eq.~\eqref{eq:pos} is valid even in the massless limit which we focus on later.

\section{Supersymmetric positivity bounds}\label{sec: supersymmetric positivity}

\subsection{Supersymmetric $\delta$-function constraint for $\mathcal N=1$ four-point amplitudes}\label{sec: supersymmetric 4-pt}

Our goal in this work is to explore the positivity bounds when Lorentz symmetry is promoted to $\mathcal N=1$ supersymmetry. Specifically, we focus on four-point interactions. At both tree and loop levels, the on-shell four-point amplitude in $\mathcal N=1$ supersymmetric theories takes the general form~\cite{Delgado:2022bho,Delgado:2023ogc,Herderschee:2019ofc,Delgado:2025oev}
\begin{equation}
\mathcal A_4 = \delta^{(4)}\left(P\right) 
\mathcal M\left(\{\eta_i\},\{\lambda_i,\tilde\lambda_i\}\right)\,,
\end{equation}
where $\lambda_i$ and $\tilde\lambda_i$ are the spinor-helicity variables associated with the external momenta
$p_i^{\alpha\dot\alpha}=\lambda_i^{\alpha}\tilde\lambda_i^{\dot\alpha}$\footnote{We provide a brief introduction of the spinor-helicity formalism in Appendix~\ref{app:spinor-helicity}.}. The Grassmann variables $\eta_i$ label which component state of the on-shell  supermultiplet is realized on the $i$-th external leg. The factor $\delta^{(4)}(P)$ enforces total momentum conservation, while $\mathcal M\left(\{\eta_i\},\{\lambda_i,\tilde\lambda_i\}\right)$ is a Lorentz scalar built from the Grassmann and spinor-helicity variables $\eta_i,\lambda_i,\tilde\lambda_i$. 

In the on-shell formalism the supersymmetry generators act on the amplitude as\footnote{Throughout this work, we adopt the $\eta$-basis~\cite{Herderschee:2019ofc,Delgado:2025oev}.}
\begin{equation}
Q^\alpha = \sum_{i=1}^{4}[i|^\alpha \frac{\partial}{\partial\eta_{i}}\,,\qquad\tilde Q^{\dot\alpha} =
\sum_{i=1}^{4}|i\rangle^{\dot\alpha} \eta_{i}\,,
\end{equation}
with $\alpha=1,2$ and $\dot\alpha=\dot 1,\dot 2$. For notational convenience, we identify $[i|^\alpha\equiv \lambda_i^\alpha$ and $  |i\rangle^{\dot\alpha}\equiv\tilde\lambda_i^{\dot{\alpha}}$. Supersymmetry invariance implies that supercharges annihilate superamplitudes, i.e.  supersymmetric Ward identities
\begin{equation}
Q^\alpha \mathcal A_4 = 0\,,\qquad\tilde Q^{\dot\alpha} \mathcal A_4 = 0\,.
\end{equation}
Motivated by the second condition, we adopt the following factorized ansatz as our starting point:
\begin{equation}
\mathcal A_4 = \delta^{(4)}(P)\delta^{(2)}(\tilde Q)F\left(\{\lambda_i,\tilde\lambda_i\}\right)\,,
\label{eq:four-point-ansatz}
\end{equation}
where $\delta^{(2)}(\tilde Q)$ is the supermomentum Grassmann delta-function defined as 
\begin{equation}
\delta^{(2)}(\tilde Q)\equiv \sum_{a<b}\langle a b\rangle \eta_a\eta_b\,,
\end{equation}
and $F$ is a purely bosonic function determined by the kinematics. Since the four-point on-shell superamplitude contains only terms quadratic~\cite{Delgado:2022bho,Delgado:2023ogc,Herderschee:2019ofc,Delgado:2025oev} in the Grassmann variables $\eta_i$, the relevant Grassmann structure is completely encoded into the $\delta^{(2)}(\tilde Q)$ factor.

With this ansatz, the second Ward identity
$\tilde Q^{\dot\alpha} \mathcal A_4 = 0$
is trivially satisfied because $\tilde Q^{\dot\alpha}\delta^{(2)}(\tilde Q)=0$ (equivalently $\tilde Q^{\dot\alpha} \tilde Q^{\dot1}\tilde Q^{\dot2}=0$ by Grassmann nilpotency).
We therefore focus on the first identity
\begin{equation}
Q^\alpha \mathcal A_4 = 0\,,
\end{equation}
which imposes a nontrivial constraint on $\delta^{(2)}(\tilde Q)$ together with momentum conservation. $F\left(\{\lambda_i,\tilde\lambda_i\}\right)$ is $\eta-$independent and therefore is not affected by the action of $Q$.

To this end we first compute the derivative of $\delta^{(2)}(\tilde  Q)$ with respect to $\eta_i$.
Starting from the definition,
\begin{equation}
\frac{\partial}{\partial\eta_i}\delta^{(2)}(\tilde Q)=\frac{\partial}{\partial\eta_i}\sum_{a<b}\langle a b\rangle \eta_a\eta_b=\sum_{a<b}\langle a b\rangle\left(\delta_{a i} \eta_b-\delta_{b i} \eta_a\right)\,,
\end{equation}
where $\delta_{ai}$ is the Kronecker delta. Rearranging the sum yields the compact form
\begin{equation}
\frac{\partial}{\partial\eta_i}\delta^{(2)}(\tilde Q)
=
\sum_{k\neq i}\langle i k\rangle \eta_k\,.
\end{equation}
Acting with $Q^\alpha$ we obtain
\begin{equation}
Q^\alpha \delta^{(2)}(\tilde Q)=\sum_{i=1}^4[i|^\alpha\frac{\partial}{\partial\eta_i}\delta^{(2)}(\tilde Q)=\sum_{i,k}[i|^\alpha\langle i k\rangle \eta_k=\sum_{k}\Big(\sum_{i}[i|^\alpha\langle i k\rangle\Big)\eta_k\,.
\end{equation}
Using momentum conservation, Eq.~\eqref{eq:momentum_conservation_spinor}, the inner sum vanishes identically, giving
\begin{equation}
Q^\alpha \delta^{(2)}(\tilde Q) = 0\,.
\end{equation}
Combining these results, we conclude that upon adopting $\delta^{(2)}(\tilde Q)$ as the given Grassmann structure, both supersymmetric Ward identities are satisfied. Thus, the ansatz Eq.~\eqref{eq:four-point-ansatz} is consistent with both supersymmetric Ward identities at four points. 

\subsection{Forward limit, crossing symmetry, and the bosonic prefactor of $\delta^{(2)}(\tilde Q)$}
\label{sec: supersymmetric forward limit}

We now examine the behavior of the supermomentum delta-function in the forward limit. This analysis shows that, upon imposing $t\to0$ with a definite holomorphic branch of the spinors, the angle-bracket coefficients in \(\delta^{(2)}(\tilde Q)=\sum_{i<j}\langle ij\rangle \eta_i\eta_j\)
reduce to a single bosonic factor. As a result, the four-point amplitude factorizes as
\(\mathcal A_4=\delta^{(4)}(P)\mathcal G(\eta_i) \tilde F(s)\), where \(\tilde F(s)\) is a purely bosonic function of the Mandelstam variable $s$. In the forward limit $t\to0$ one may choose the holomorphic solution in which the angle brackets connecting particles $(1,3)$ and $(2,4)$ vanish,
\begin{equation}
\langle 1 3\rangle=\langle 2 4\rangle=0 \qquad (\text{as } t\to0)\,.
\end{equation}
Since the angle brackets vanish, the corresponding holomorphic spinors become linearly dependent, implying
\begin{equation}
|3\rangle=a |1\rangle\,,\qquad
|4\rangle=b |2\rangle\,,
\end{equation}
for some nonzero complex constants $a$ and $b$.

With these kinematic simplifications the supermomentum delta-function reduces to the four nonzero terms. Because $|3\rangle$ and $|4\rangle$ are proportional to $|1\rangle$ and $|2\rangle$ in the forward limit, respectively, each remaining angle bracket is proportional to the common factor $\langle12\rangle$. Factoring this out and we obtain
\begin{equation}
\begin{aligned}
\delta^{(2)}(\tilde Q)\Big|_{t\to0}
&=
\langle 1 2\rangle \eta_1\eta_2
+\langle 1 3\rangle \eta_1\eta_3
+\langle 1 4\rangle \eta_1\eta_4
+\langle 2 3\rangle \eta_2\eta_3
+\langle 2 4\rangle \eta_2\eta_4
+\langle 3 4\rangle \eta_3\eta_4 \\
&=
\langle 1 2\rangle \eta_1\eta_2
+\langle 1 4\rangle \eta_1\eta_4
+\langle 2 3\rangle \eta_2\eta_3
+\langle 3 4\rangle \eta_3\eta_4\\
&=
\langle 1 2\rangle \eta_1\eta_2
+b\langle 1 2\rangle \eta_1\eta_4
-a\langle 1 2\rangle \eta_2\eta_3
+ab\langle 1 2\rangle \eta_3\eta_4\\
&=\langle 1 2\rangle \mathcal{G}(\eta_i)\,,
\end{aligned}
\end{equation}
where \(\mathcal G(\eta_i)\) is a purely numerical linear combination of the \(\eta_i\eta_j\) monomials (independent of the spinor-helicity variables). This demonstrates that, in the forward limit, the bosonic part of \(\delta^{(2)}(\tilde Q)\) reduces to a single independent factor which can be absorbed into the overall bosonic dynamical factor \(F\).

Finally, absorbing the common factor \(\langle1 2\rangle\) into \(F\), we arrive at the desired factorization of the full amplitude in the forward, crossing-symmetric kinematics:
\begin{equation}
\mathcal A_4=\delta^{(4)}(P)\delta^{(2)}(\tilde Q)F(\{\lambda_i,\tilde\lambda_i\})
\xrightarrow[\text{crossing symmetry}]{\text{forward limit}}\delta^{(4)}(P) \mathcal G(\eta_i) \tilde F(s)\,.
\label{eq:forward_limit_superamplitude}
\end{equation}
with \(\tilde F(s)\) a symmetric (under the imposed crossing) purely bosonic function of the Mandelstam variable $s$.

\subsection{Positivity of supersymmetric amplitudes}\label{sec: positivity of  supersymmetry}

With Eq.~\eqref{eq:forward_limit_superamplitude} established, we now turn to the positivity of supersymmetric amplitudes. The key observation is that, once the supersymmetric Ward identities have been imposed, the Grassmann dependence $\mathcal G(\eta_i)$ becomes universal and decouples from the analytic structure in the Mandelstam variable $s$. Consequently, the analysis of positivity can be carried out solely at the level of $\tilde F(s)$, which inherits the same analyticity, crossing symmetry, and boundedness properties as a standard non-supersymmetric forward scattering amplitude.
Dispersion-relation arguments then lead directly to the positivity constraints on the low-energy expansion coefficients of $\tilde F(s)$. 

As a concrete illustration, consider a four-point superamplitude of $\mathcal N=1$ chiral superfields. Once the dependence on the Grassmann variables $\eta_i$ has been completely factorized from the kinematic dependence on $s$, the remaining bosonic function $\tilde F(s)$ maps to an ordinary forward scattering amplitude,
\begin{equation}
\tilde F(s) = \mathcal A^{ab\to ab}(s)\,,
\end{equation}
where $\tilde F(s)$ coincides with a specified physical elastic amplitude. The reference channel may involve the scalar $\phi$ or the fermion $\psi$ components of the chiral multiplet, provided this identification is consistent with the corresponding component amplitude.\footnote{Auxiliary fields are eliminated algebraically and do not appear as external states.} 

For example, let us focus on the pure scalar channel $\mathcal A^{\phi\phi\to\phi\phi}(s)$ and make the identification
\begin{equation}
\tilde F(s)=\mathcal A^{\phi\phi\to\phi\phi}(s).
\end{equation}
In the forward limit, the dim-8 coefficient of this component amplitude satisfies the positivity bound~\eqref{eq:pos}. The identification therefore imposes the same constraint on the corresponding coefficient of $\tilde F(s)$. Since $\tilde F(s)$ is the common prefactor of the forward superamplitude under consideration, this scalar positivity bound constrains the full supersymmetric interaction. In this sense, supersymmetry packages the positivity of individual component amplitudes into a single superamplitude-level constraint.

It is worth emphasizing that this conclusion relies crucially on the complete factorization of the Grassmann variables $\eta_i$ from the kinematic dependence on $s$ in the forward limit, a property of $\mathcal{N}=1$ four-point amplitudes. For higher-point processes, neither the forward limit nor crossing symmetry is uniquely defined in a form suitable for standard dispersion-relation arguments. Positivity bounds of the type derived here do not generalize straightforwardly beyond the four-point case.

In addition, in the massive case, the Grassmann dependence cannot be fully disentangled from the $s$-dependence in the forward limit. Specifically, the $s$-dependence required by the supersymmetric constraint $\delta^{(2)}(\tilde Q)$ does not separate completely. Nevertheless, this does not preclude the possibility that there exists a representation of the $\lambda$ variables in which the resulting relations take a simpler form. In such a representation, the positivity conditions for the individual amplitude components can be checked more readily.

\section{Examples of supersymmetry-related operator coefficients subject to positivity constraints}\label{sec: examples}
\subsection{Single chiral multiplet}
\label{sec:single}
We begin by analyzing the $2\to2$ scattering amplitudes of a single $\mathcal{N}=1$ chiral supermultiplet, which contains a complex scalar $\phi$ and its fermionic partner $\psi$. Specifically, we focus on three distinct processes: pure-scalar ($\phi\phi\rightarrow\phi\phi$), mixed scalar--fermion ($\phi\psi\rightarrow\phi\psi$), and pure-fermion ($\psi\psi\rightarrow\psi\psi$) scattering. Our analysis proceeds in three steps: (i) we identify the independent operators contributing to the relevant contact interactions at tree level; (ii), we compute the scattering amplitudes associated with each of these operators; (iii), we impose supersymmetric Ward identities to derive exact relations among the component amplitudes and their associated Wilson coefficients. We then demonstrate that these supersymmetry-enforced relations are fully consistent with the positivity bounds derived in the non-supersymmetric framework.

\begin{figure}[!t]
\centering

\begin{subfigure}{0.32\textwidth}
\centering
\begin{fmffile}{feyn1}
\fmfframe(2,8)(2,2){%
\begin{fmfgraph*}(100,100)
    \fmfleft{i1,i2}
    \fmfright{o1,o2}
    \fmf{scalar}{i1,w1}
    \fmf{scalar}{i2,w1}
    \fmf{scalar}{w1,o1}
    \fmf{scalar}{w1,o2}
    \fmfv{lab=$\phi(p_1)$,lab.dist=-0.05w}{i2}
    \fmfv{lab=$\phi(p_2)$,lab.dist=-0.05w}{i1}
    \fmfv{lab=$\phi(p_3)$,lab.dist=-0.05w}{o2}
    \fmfv{lab=$\phi(p_4)$,lab.dist=-0.05w}{o1}
\end{fmfgraph*}%
}
\end{fmffile}
\caption{$\phi\phi\rightarrow\phi\phi$}
\label{fig:contact-diagrams1a}
\end{subfigure}
\hfill
\begin{subfigure}{0.32\textwidth}
\centering
\begin{fmffile}{feyn2}
\fmfframe(2,8)(2,2){%
\begin{fmfgraph*}(100,100)
    \fmfleft{i1,i2}
    \fmfright{o1,o2}
    \fmf{fermion}{i1,w1}
    \fmf{scalar}{i2,w1}
    \fmf{fermion}{w1,o1}
    \fmf{scalar}{w1,o2}
    \fmfv{lab=$\phi(p_1)$,lab.dist=-0.05w}{i2}
    \fmfv{lab=$\psi(p_2)$,lab.dist=-0.05w}{i1}
    \fmfv{lab=$\phi(p_3)$,lab.dist=-0.05w}{o2}
    \fmfv{lab=$\psi(p_4)$,lab.dist=-0.05w}{o1}
\end{fmfgraph*}%
}
\end{fmffile}
\caption{$\phi\psi\rightarrow\phi\psi$}
\label{fig:contact-diagrams1b}
\end{subfigure}
\hfill
\begin{subfigure}{0.32\textwidth}
\centering
\begin{fmffile}{feyn3}
\fmfframe(2,8)(2,2){%
\begin{fmfgraph*}(100,100)
    \fmfleft{i1,i2}
    \fmfright{o1,o2}
    \fmf{fermion}{i1,w1}
    \fmf{fermion}{i2,w1}
    \fmf{fermion}{w1,o1}
    \fmf{fermion}{w1,o2}
    \fmfv{lab=$\psi(p_1)$,lab.dist=-0.05w}{i2}
    \fmfv{lab=$\psi(p_2)$,lab.dist=-0.05w}{i1}
    \fmfv{lab=$\psi(p_3)$,lab.dist=-0.05w}{o2}
    \fmfv{lab=$\psi(p_4)$,lab.dist=-0.05w}{o1}
\end{fmfgraph*}%
}
\end{fmffile}
\caption{$\psi\psi\rightarrow\psi\psi$}
\label{fig:contact-diagrams1c}
\end{subfigure}

\caption{Four-point contact interaction diagrams for the component scattering processes in the chiral multiplet.}
\label{fig:contact-diagrams1}
\end{figure}
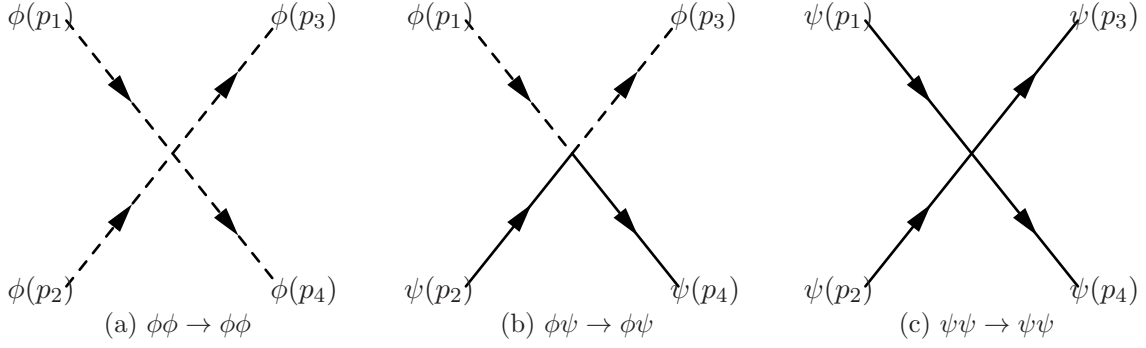

\paragraph{Scalar--scalar scattering.}
We consider a contact four-point vertex for the scattering process 
$\phi \phi \to \phi \phi$, where $\phi$ denotes the scalar component of the chiral superfield 
$\Phi(x,\theta,\bar\theta)\big|_{\theta=\bar\theta=0}=\phi(x)$. 
This channel corresponds to the scattering of two scalars $\phi$ into two scalars $\phi$, as illustrated in Fig.~\ref{fig:contact-diagrams1}\subref{fig:contact-diagrams1a}.

At mass-dimension eight (dim-8), the independent four-point operator with two $\phi^\dagger$, two $\phi$ fields and four derivatives are chosen to be \footnote{For notational simplicity, we suppress repeated initial/final state labels henceforth, writing
\(c^{ab}\equiv c^{ab\to ab}\) and
\(\mathcal{O}^{ab}\equiv\mathcal{O}^{ab\to ab}\),
unless stated otherwise.}: 

\begin{equation}
\mathcal{O}^{\phi \phi }_1 
= 
\big(\partial_\mu \phi^\dagger   \partial^\mu \phi \big)
\big(\partial_\nu \phi^\dagger   \partial^\nu \phi \big)\,,
\label{eq:O1}
\end{equation}
\begin{equation}
\mathcal{O}^{\phi \phi }_2 
= 
\big(\partial_\mu \phi^\dagger   \partial_\nu \phi \big)
\big(\partial^\mu \phi^\dagger   \partial^\nu \phi \big)\,.
\label{eq:O2}
\end{equation}
The subscript distinguishes different independent operator structures within the same process. The tree-level $\phi \phi \to \phi \phi$ amplitude is\footnote{Through out this paper, the convention of Wilson coefficients  is chosen such that $\mathcal{L}_{\rm dim-8} = \underset{i}{\sum} \frac{c_i}{\Lambda^4} \mathcal{O}_i $, where the coefficient $c_i$ has the same labels as the corresponding $\mathcal{O}_i$ ({\it e.g.}, $c^{\phi\phi}_1$ is the coefficient of $\mathcal{O}^{\phi \phi }_1$).  }
\begin{equation}
\mathcal{A}(\phi^\dagger,\phi^\dagger,\phi,\phi)= \frac{c^{\phi \phi }_1}{\Lambda^4} \left( \frac{t^2}{2} + \frac{u^2}{2} \right)+\frac{c^{\phi \phi }_2}{\Lambda^4}s^2\,.
\end{equation}
The forward-limit positivity bound therefore yields:
\begin{equation}
c^{\phi\phi}_{2}+\frac{1}{2} c^{\phi\phi}_{1}>0\,.
\label{pos1}
\end{equation}

\paragraph{Mixed scalar-fermion scattering.}
We now turn to the $\phi \psi \to \phi \psi$ channel, where $\psi$ denotes the fermionic component of the chiral superfield. The relevant contact interaction is depicted in Fig.~\ref{fig:contact-diagrams1}\subref{fig:contact-diagrams1b}. We choose the independent dim-8 four-point operators with two fermions and two scalars carrying three derivatives to be
\begin{equation}
\mathcal{O}^{\phi \psi }_1 
= i
\big(\bar\psi \gamma^\nu\partial^\mu\psi\big) 
\big(\phi^\dagger \partial_{\mu}\partial_{\nu}\phi\big)\,,
\label{eq:O1_fermion}
\end{equation}
\begin{equation}
\mathcal{O}^{\phi \psi }_2 
= i
\big(\bar\psi \gamma^\nu\partial^\mu\psi\big) 
\big(\partial_{\mu}\partial_{\nu} \phi^\dagger \phi\big)\,,
\label{eq:O2_fermion}
\end{equation}
Computing the tree-level four-point amplitude 
\(\mathcal{A}(\phi^\dagger,\bar\psi,\phi,\psi)\) 
in the all-outgoing convention, with external momenta 
\(p_1, p_2, p_3, p_4\) assigned to \(\phi^\dagger,\bar\psi,\phi,\psi\), respectively, we find
\begin{equation}
\mathcal{A}(\phi^\dagger,\bar\psi,\phi,\psi)
= -\frac{c^{\phi \psi }_1}{2\Lambda^4} s \langle 4 1\rangle [1 2]
+\frac{c^{\phi \psi }_2}{2\Lambda^4} u \langle 4 1\rangle [1 2]\,.
\end{equation}
The corresponding forward-limit positivity bound is therefore given by
\begin{equation}
c^{\phi\psi}_{2}+c^{\phi\psi}_{1}<0\,.
\label{pos2}
\end{equation}

\paragraph{Pure-fermion scattering.}
Extending the analysis to \(\psi \psi \to \psi \psi\) (Fig.~\ref{fig:contact-diagrams1}\subref{fig:contact-diagrams1c}), the independent dim-8 operator with four fermions and two derivatives is
\begin{equation}
\mathcal{O}^{\psi \psi } 
= 
\big(\bar\psi \gamma^\mu\overset{\leftrightarrow}{\partial^\nu}\psi\big)
\big(\bar\psi \gamma_\mu\overset{\leftrightarrow}{\partial_\nu}\psi\big)\,,
\label{eq:O3_fermion}
\end{equation}
where \(\overset{\leftrightarrow}{\partial^\nu} = \partial^\nu - \overset{\leftarrow}{\partial^\nu}\). The amplitude therefore evaluates to
\begin{equation}
\mathcal{A}_{\mathcal{O}^{\psi \psi }}(\bar\psi,\bar\psi,\psi,\psi)
= -4 \frac{c^{\psi \psi }}{\Lambda^4} (2s+t+u) [1 2]\langle 3 4\rangle
= -4 \frac{c^{\psi \psi }}{\Lambda^4} s [1 2]\langle 3 4\rangle\,,
\end{equation}
and the corresponding forward-limit positivity bound is given by
\begin{equation}
c^{\psi \psi}>0\,.
\label{pos3}
\end{equation}

These results complete the enumeration of independent contact operators and their tree-level amplitudes for the scalar, mixed, and purely fermionic channels. They provide the basis for exploiting supersymmetric Ward identities
and for analyzing the relations among operator coefficients within the chiral superfield framework.

\paragraph{Supersymmetric constraints from Ward identities.}
We introduce the action of the supercharge on asymptotic states:
$a^{+}$ and $a^{-}$ denote creation/annihilation operators for $\phi$ and $\phi^\dagger$, while $b^{+}$ and $b^{-}$ are those for $\psi$ and $\bar{\psi}$. Using supersymmetry invariance of the vacuum, $Q^\dagger |0\rangle = 0$, we compute the vacuum matrix elements
\begin{equation}
\langle 0 | [Q^\dagger, a^+(p_1)  a^+(p_2)  a^-(p_3)  b^-(p_4)] | 0 \rangle = 0\,,
\end{equation}
\begin{equation}
\langle 0 | [Q^\dagger, a^+(p_1)  b^+(p_2)  b^-(p_3)  b^-(p_4)] | 0 \rangle = 0\,,
\end{equation}
where $Q^{\dagger}$ is the Hermitian conjugate of the supercharge associated with the chiral superfield representation. These vacuum matrix elements vanish as a consequence of $Q^\dagger |0\rangle = 0$, together with the standard (anti-)commutation relations between the supercharge and the creation and annihilation operators.

Expanding the commutators using the supersymmetry algebra and evaluating their action on one-particle states, the above identities translate into linear relations among on-shell scattering amplitudes involving different external states within the same supermultiplet. In particular, we obtain the following nontrivial relations among the four-point amplitudes (see Appendix~\ref{appendix: supersymmetric Ward Identity} for details):
\begin{equation}
\mathcal{A}
(\phi^\dagger,\phi^\dagger,\phi,\phi) = -\frac{\langle 1 2\rangle}{\langle 1 4\rangle}\mathcal{A}
(\phi^\dagger,\bar\psi,\phi,\psi) \,,
\label{wardrelation1}
\end{equation}
\begin{equation}
\mathcal{A}
(\phi^\dagger,\bar\psi,\phi,\psi) = -\frac{\langle 1 4\rangle}{\langle 4 3\rangle}\mathcal{A}(\bar\psi,\bar\psi,\psi,\psi) \,.
\label{wardrelation2}
\end{equation}
Substituting the tree-level amplitudes into the Ward identities yields linear relations among the Wilson coefficients:
\begin{equation}
\begin{aligned}
&c^{\phi\phi}_{1}=c^{\phi\psi}_{2}=0\,, \qquad
c^{\phi\phi}_{2}=-\frac{1}{2}c^{\phi\psi}_{1}=4c^{\psi\psi}\,.
\label{relation1}
\end{aligned}
\end{equation}
The vanishing coefficients multiply kinematic structures that cannot be matched across the Ward identities. Supersymmetry therefore excludes these structures from the four-point superamplitude.
Equation~\eqref{relation1} directly relates the Wilson coefficients across all three component channels and is manifestly consistent with Eqs.~\eqref{pos1}, \eqref{pos2}, and \eqref{pos3}, providing a nontrivial cross-check of the supersymmetric structure. 

Equivalently, one may obtain these relations by expanding the off-shell superspace integration in components\footnote{At mass dimension-eight, $D\Phi D\Phi\overline{D}\Phi^\dagger\overline{D}\Phi^\dagger$ is the independent operator with two chiral superfields and two anti-chiral superfields.}:
\begin{equation}
    \mathcal{L}\supset \frac{c}{\Lambda^4}\int d^4\theta\, D\Phi D\Phi\overline{D}\Phi^\dagger\overline{D}\Phi^\dagger= \frac{4c}{\Lambda^4}\mathcal{O}^{\phi\phi}_2-\frac{8c}{\Lambda^4}\mathcal{O}^{\phi\psi}_1+\frac{c}{\Lambda^4}\mathcal{O}^{\psi\psi} \,,
\end{equation}
where $D$ and $\overline D$ are super-covariant derivatives; $\Phi$ and $\Phi^\dagger$ are chiral superfields and anti-chiral superfields, respectively. We see that the relative ratio of corresponding couplings reproduces the coefficient relations in complete agreement with the Ward identity derivation. 

In this example, we see that Eq.~\eqref{eq:forward_limit_superamplitude}  is consistent with Eqs.~\eqref{wardrelation1}-\eqref{relation1}, implying that component amplitudes within a single supermultiplet must satisfy the same positivity bounds. However, we emphasize that the inverse statement, the bound of a superamplitude is completely determined by one of the component bound, does not necessarily holds. Specifically, the lack of a bound on a component amplitude does not imply that the Wilson coefficient is still unconstrained after supersymmetrization. We will discuss a concrete example in the next section.

Finally, the numbers of independent Wilson coefficients contributing to the scattering amplitudes considered in the single-flavor case, with and without supersymmetry, are summarized in Table~\ref{tab:susy_operators1}.
\begin{table}[!t]
\centering
\begin{tabular}{|c|c|c|}
\hline
Scattering amplitude & Without SUSY & With SUSY \\
\hline
$\mathcal{A}
(\phi^\dagger,\phi^\dagger,\phi,\phi)$ & 2 & 1 \\
\hline
$\mathcal{A}
(\phi^\dagger,\bar\psi,\phi,\psi)$ & 2 & 1 \\
\hline
$\mathcal{A}(\bar\psi,\bar\psi,\psi,\psi)$ & 1 &  1\\
\hline
$\text{all}$ & 5 &  1\\
\hline
\end{tabular}
\caption{Number of independent dim-8 operators with or without SUSY for the single-chiral-multiplet scenario.}
\label{tab:susy_operators1}
\end{table}

\subsection{Two(-flavor) chiral supermultiplets }
\begin{figure}[!t]
\centering

\begin{subfigure}{0.32\textwidth}
\centering
\begin{fmffile}{feyn4}
\fmfframe(2,8)(2,2){%
\begin{fmfgraph*}(100,100)
    \fmfleft{i1,i2}
    \fmfright{o1,o2}
    \fmf{scalar}{i1,w1}
    \fmf{scalar}{i2,w1}
    \fmf{scalar}{w1,o1}
    \fmf{scalar}{w1,o2}
    \fmfv{lab=$\phi_1(p_1)$,lab.dist=-0.05w}{i2}
    \fmfv{lab=$\phi_2(p_2)$,lab.dist=-0.05w}{i1}
    \fmfv{lab=$\phi_1(p_3)$,lab.dist=-0.05w}{o2}
    \fmfv{lab=$\phi_2(p_4)$,lab.dist=-0.05w}{o1}
\end{fmfgraph*}%
}
\end{fmffile}
\caption{$\phi_1\phi_2\rightarrow\phi_1\phi_2$}
\label{fig:contact-diagrams3a}
\end{subfigure}
\hfill
\begin{subfigure}{0.32\textwidth}
\centering
\begin{fmffile}{feyn5}
\fmfframe(2,8)(2,2){%
\begin{fmfgraph*}(100,100)
    \fmfleft{i1,i2}
    \fmfright{o1,o2}
    \fmf{fermion}{i1,w1}
    \fmf{scalar}{i2,w1}
    \fmf{fermion}{w1,o1}
    \fmf{scalar}{w1,o2}
    \fmfv{lab=$\phi_1(p_1)$,lab.dist=-0.05w}{i2}
    \fmfv{lab=$\psi_2(p_2)$,lab.dist=-0.05w}{i1}
    \fmfv{lab=$\phi_1(p_3)$,lab.dist=-0.05w}{o2}
    \fmfv{lab=$\psi_2(p_4)$,lab.dist=-0.05w}{o1}
\end{fmfgraph*}%
}
\end{fmffile}
\caption{$\phi_1\psi_2\rightarrow\phi_1\psi_2$}
\label{fig:contact-diagrams3b}
\end{subfigure}
\hfill
\begin{subfigure}{0.32\textwidth}
\centering
\begin{fmffile}{feyn6}
\fmfframe(2,8)(2,2){%
\begin{fmfgraph*}(100,100)
    \fmfleft{i1,i2}
    \fmfright{o1,o2}
    \fmf{fermion}{i1,w1}
    \fmf{fermion}{i2,w1}
    \fmf{fermion}{w1,o1}
    \fmf{fermion}{w1,o2}
    \fmfv{lab=$\psi_1(p_1)$,lab.dist=-0.05w}{i2}
    \fmfv{lab=$\psi_2(p_2)$,lab.dist=-0.05w}{i1}
    \fmfv{lab=$\psi_1(p_3)$,lab.dist=-0.05w}{o2}
    \fmfv{lab=$\psi_2(p_4)$,lab.dist=-0.05w}{o1}
\end{fmfgraph*}%
}
\end{fmffile}
\caption{$\psi_1\psi_2\rightarrow\psi_1\psi_2$}
\label{fig:contact-diagrams3c}
\end{subfigure}

\caption{Four-point contact interaction diagrams for the component scattering processes involving two distinct chiral multiplets.}
\label{fig:contact-diagrams3}
\end{figure}
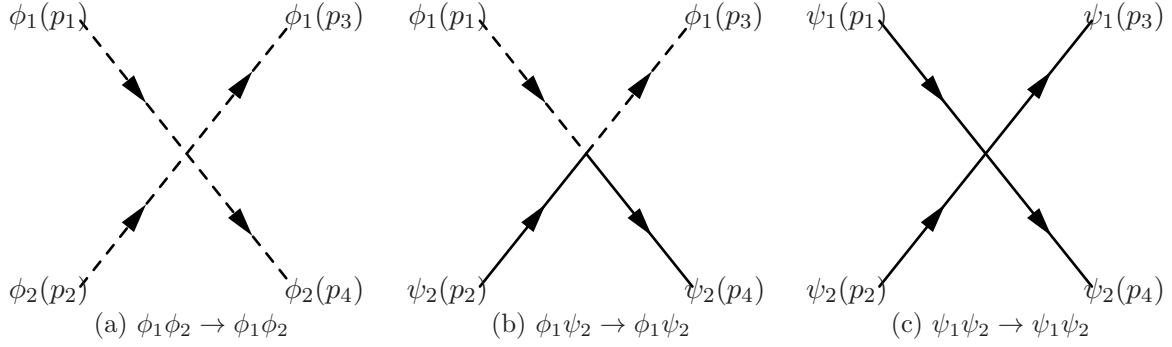

We now consider two-flavor chiral supermultiplets and examine the corresponding scattering processes. For simplicity, we assume flavor is conserved, and focus only on the amplitudes involving two different flavors (since the results for single-flavor amplitudes would be the same as in Sec.~\ref{sec:single}). In this setup, we find that a minimal set of representative component scattering processes is sufficient to illustrate how supersymmetry relates several positivity bounds through the supersymmetric Ward identities. Moreover, we find an example of an amplitude that would otherwise be free from any positivity constraint, but is in fact subject to a nontrivial supersymmetry-enforced restriction.\footnote{For details on how this example is obtained, see Appendix~\ref{appendix: supersymmetric Ward Identity}.} The relevant Feynman diagrams for the elastic scattering processes are shown in Fig.~\ref{fig:contact-diagrams3}, while the corresponding independent dim-8 operators and the resulting amplitudes are summarized in Table~\ref{tab:twochiraloperators}.

\begin{table}[!t]
\centering
\footnotesize

\begin{tblr}{
width=\textwidth,
colspec={|Q[c,m,0.07]|Q[c,m,0.49]|Q[c,m,0.49]|},
hlines,
vlines,
row{1}={font=\bfseries},
}

&
Operators
&
Amplitudes
\\

$\phi_1\phi_2\rightarrow\phi_1\phi_2$

&
{$
\begin{aligned}
\mathcal O^{\phi_1\phi_2}_1
&=
(\partial_\mu\phi_1^\dagger\partial^\mu\phi_1)
(\partial_\nu\phi_2^\dagger\partial^\nu\phi_2)\,,
\\[0.8em]
\mathcal O^{\phi_1\phi_2}_2
&=
(\partial_\mu\phi_1^\dagger\partial^\mu\phi_2^\dagger)
(\partial_\nu\phi_1\partial^\nu\phi_2)\,,
\\[0.8em]
\mathcal O^{\phi_1\phi_2}_3
&=
(\partial_\mu\phi_1^\dagger\partial^\mu\phi_2)
(\partial_\nu\phi_2^\dagger\partial^\nu\phi_1)\,.
\end{aligned}
$}

&

{$
\begin{aligned}
\mathcal A(\phi_1^\dagger,\phi_2^\dagger,\phi_1,\phi_2)
&=
\frac{c^{\phi_1\phi_2}_1}{\Lambda^4}\frac{t^2}{4}
+\frac{c^{\phi_1\phi_2}_2}{\Lambda^4}\frac{s^2}{4} \\
&+\frac{c^{\phi_1\phi_2}_3}{\Lambda^4}\frac{u^2}{4}\,.
\end{aligned}
$}

\\

$\phi_1\psi_2\rightarrow\phi_1\psi_2$

&

{$
\begin{aligned}
\mathcal O^{\phi_1\psi_2}_1
&=i
(\bar\psi_2\gamma^\nu\partial^\mu\psi_2)
(\phi_1^\dagger\partial_{\mu}\partial_{\nu}\phi_1)\,,
\\[0.8em]
\mathcal O^{\phi_1\psi_2}_2
&=i
(\bar\psi_2\gamma^\nu\partial^\mu\psi_2)
(\partial_{\mu}\partial_{\nu}\phi_1^\dagger\phi_1)\,.
\end{aligned}
$}

&

{$
\begin{aligned}
\mathcal A(\phi_1^\dagger,\bar\psi_2,\phi_1,\psi_2)
&=
-\frac{c^{\phi_1\psi_2}_1}{2\Lambda^4}s\langle41\rangle[12] \\
&+\frac{c^{\phi_1\psi_2}_2}{2\Lambda^4}u\langle41\rangle[12]\,.
\end{aligned}
$}

\\

$\psi_1\psi_2\rightarrow\psi_1\psi_2$

&

{$
\begin{aligned}
\mathcal O^{\psi_1\psi_2}_1
&=
(\bar\psi_1\gamma^\mu
\overleftrightarrow{\partial^\nu}\psi_1)
(\bar\psi_2\gamma_\mu
\overleftrightarrow{\partial_\nu}\psi_2)\,,
\\[0.8em]
\mathcal O^{\psi_1\psi_2}_2
&=
(\bar\psi_1\gamma^\mu
\overleftrightarrow{\partial^\nu}\psi_2)
(\bar\psi_2\gamma_\mu
\overleftrightarrow{\partial_\nu}\psi_1)\,.
\end{aligned}
$}

&

{$
\begin{aligned}
\mathcal A(\bar\psi_1,\bar\psi_2,\psi_1,\psi_2)
&=
-2\frac{c^{\psi_1\psi_2}_1}{\Lambda^4}(s-u)[12]\langle34\rangle \\
&-2\frac{c^{\psi_1\psi_2}_2}{\Lambda^4}(s-t)[12]\langle34\rangle\,.
\end{aligned}
$}

\\

\end{tblr}

\caption{Dim-8 operators and the corresponding four-point amplitudes for the component scattering processes in the two-chiral-multiplet theory (where \(\overset{\leftrightarrow}{\partial^\nu} = \partial^\nu - \overset{\leftarrow}{\partial^\nu}\)).}
\label{tab:twochiraloperators}

\end{table}

Following a similar derivation, we obtain the corresponding forward-limit positivity bounds, which are
\begin{equation}
2c^{\psi_1 \psi_2 }_{1}+c^{\psi_1 \psi_2 }_{2} > 0\,,\ \ 
c^{\phi_1 \psi_2 }_{1}
+ c^{\phi_1 \psi_2 }_{2} < 0\,,\ \ 
c^{\phi_1 \phi_2 }_{2}
+ c^{\phi_1 \phi_2 }_{3} > 0\,.
\label{pospsi}
\end{equation}
Supersymmetric Ward identities relate the three channels in the following way: 
\begin{equation}
\mathcal{A}
(\phi^\dagger_1,\phi^\dagger_2,\phi_1,\phi_2) = -\frac{\langle 1 2\rangle}{\langle 1 4\rangle}\mathcal{A}
(\phi^\dagger_1,\bar\psi_2,\phi_1,\psi_2)\,,
\end{equation}
\begin{equation}
\mathcal{A}
(\phi^\dagger_1,\bar\psi_2,\phi_1,\psi_2) = -\frac{\langle 1 4\rangle}{\langle 4 3\rangle}\mathcal{A}(\bar\psi_1,\bar\psi_2,\psi_1,\psi_2)\,.
\end{equation}
Inserting the explicit amplitudes from Table~\ref{tab:twochiraloperators}, we obtain the following relations among Wilson coefficients:
\begin{equation}
\begin{aligned}
&c^{\phi_1 \phi_2 }_{3}=-c^{\phi_1 \phi_2 }_{1}\,, \qquad
c^{\phi_1 \psi_2 }_{2}=c^{\phi_1 \phi_2 }_{1}\,, \qquad
c^{\phi_1 \psi_2 }_{1}=-\frac{c^{\phi_1 \phi_2 }_{1}+c^{\phi_1 \phi_2 }_{2}}{2}\,, \\
&c^{\psi_1 \psi_2 }_{1}=\frac{-3c^{\phi_1 \phi_2 }_{1}+c^{\phi_1 \phi_2 }_{2}}{24}\,, \qquad
c^{\psi_1 \psi_2 }_{2}=\frac{3c^{\phi_1 \phi_2 }_{1}+c^{\phi_1 \phi_2 }_{2}}{24}\,.
\end{aligned}
\label{eq: two chiral relations}
\end{equation}

The Ward identities relate the $t^2$ and $u^2$ kinematic structures in the scalar amplitude. Their relative sign is fixed by the spinor-bracket factors, yielding $c^{\phi_1 \phi_2 }_{3}=-c^{\phi_1 \phi_2 }_{1}$. Taking $c^{\phi_1 \phi_2 }_{1}$ and $c^{\phi_1 \phi_2 }_{2}$ as independent coefficients, all remaining coefficients are determined by Eq.~\eqref{eq: two chiral relations}. They are manifestly consistent with the positivity bounds Eq.~\eqref{pospsi}, providing a direct verification that the supersymmetric relations automatically respect the required positivity conditions.

The numbers of independent Wilson coefficients contributing to the scattering amplitudes for the two-chiral-multiplet scenario are summarized in Table~\ref{tab:susy_operators2}.
\begin{table}[!t]
\centering
\begin{tabular}{|c|c|c|}
\hline
Number of independent parameters & Without SUSY & With SUSY \\
\hline
$\mathcal{A}
(\phi^\dagger_1,\phi^\dagger_2,\phi_1,\phi_2)$ & 3 & 2 \\
\hline
$\mathcal{A}
(\phi^\dagger_1,\bar\psi_2,\phi_1,\psi_2)$ & 2 & 2 \\
\hline
$\mathcal{A}(\bar\psi_1,\bar\psi_2,\psi_1,\psi_2)$ & 2 &  2\\
\hline
$\text{all}$ & 7 &  2\\
\hline
\end{tabular}
\caption{Number of independent dim-8 operators with or without SUSY for the two-chiral-multiplet scenario.}
\label{tab:susy_operators2}
\end{table}

\paragraph{An `exceptional' example}
We now consider the following example. Unlike previous examples, this amplitude is identified directly from the supersymmetric Ward identity,
\begin{equation}
\mathcal{A}
(\phi^\dagger_1,\phi_2,\psi_1,\bar\psi_2)
=
\frac{\langle 1 3\rangle}{\langle 1 4\rangle}
\mathcal{A}
(\phi^\dagger_1,\phi_2,\phi_1,\phi^\dagger_2)\,.
\end{equation}
The amplitude on the right-hand side is generated by the operators in the first row of Table~\ref{tab:twochiraloperators}.  Compared with the scattering process considered earlier, the external particles labeled by $2$ and $4$ are interchanged. Consequently, the corresponding amplitude is obtained by exchanging the Mandelstam variables $s$ and $u$:
\begin{equation}
\mathcal{A}(\phi^\dagger_1,\phi_2,\phi_1,\phi^\dagger_2)
= \frac{c^{\phi_1 \phi_2 }_1}{\Lambda^4}\left( \frac{t^2}{4} \right)
+\frac{c^{\phi_1 \phi_2 }_2}{\Lambda^4}\left( \frac{u^2}{4} \right)
+\frac{c^{\phi_1 \phi_2 }_3}{\Lambda^4}\left( \frac{s^2}{4} \right)\,.
\end{equation}

\begin{figure}[!t]
\centering
\begin{fmffile}{feyn9}
\fmfframe(2,8)(2,2){%
\begin{fmfgraph*}(100,100)
    \fmfleft{i1,i2}
    \fmfright{o1,o2}
    \fmf{scalar}{i1,w1}
    \fmf{scalar}{i2,w1}
    \fmf{fermion}{w1,o1}
    \fmf{fermion}{w1,o2}
    \fmfv{lab=$\phi_1(p_1)$,lab.dist=-0.05w}{i2}
    \fmfv{lab=$\phi^\dagger_2(p_2)$,lab.dist=-0.05w}{i1}
    \fmfv{lab=$\psi_1(p_3)$,lab.dist=-0.05w}{o2}
    \fmfv{lab=$\bar\psi_2(p_4)$,lab.dist=-0.05w}{o1}
\end{fmfgraph*}%
}
\end{fmffile}
\caption{Contact diagram for \(\phi_1\phi^\dagger_2 \to \psi_1\bar\psi_2\) scattering.}
\label{fig:contact-diagram9}
\end{figure}
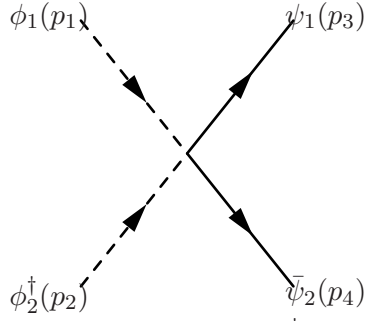

The Feynman diagram corresponding to the amplitude on the left-hand side is shown in Fig.~\ref{fig:contact-diagram9}. Meanwhile, we consider the following independent dimension-eight operators,

\begin{equation}
\mathcal{O}^{\phi_1 \phi^\dagger_2 \to \psi_1 \bar\psi_2}_1 
= i
\big(\bar\psi_2 \gamma^\nu\partial^\mu\psi_1\big) 
\big(\phi^\dagger_1 \partial_{\mu}\partial_{\nu}\phi_2\big)\,,
\end{equation}
\begin{equation}
\mathcal{O}^{\phi_1 \phi^\dagger_2 \to \psi_1 \bar\psi_2}_2 
= i
\big(\bar\psi_2 \gamma^\nu\partial^\mu\psi_1\big) 
\big(\partial_{\mu}\partial_{\nu}\phi^\dagger_1 \phi_2\big)\,,
\end{equation}
whose combined contribution to the amplitude is
\begin{equation}
\mathcal{A}(\phi^\dagger_1,\phi_2,\psi_1,\bar\psi_2)
= -\frac{c^{\phi_1 \phi^\dagger_2 \to \psi_1 \bar\psi_2}_1}{2\Lambda^4} u \langle 3 1\rangle [1 4]
+\frac{c^{\phi_1 \phi^\dagger_2 \to \psi_1 \bar\psi_2}_2}{2\Lambda^4} t \langle 3 1\rangle [1 4]\,.
\end{equation}
Applying the Ward identity, we obtain the following relation:
\begin{equation}
\begin{aligned}
&c^{\phi_1 \phi_2 }_{3}=-c^{\phi_1 \phi_2 }_{1}\,, \qquad
c^{\phi_1 \phi^\dagger_2 \to \psi_1 \bar\psi_2}_{1}=\frac{c^{\phi_1 \phi_2 }_{2}-c^{\phi_1 \phi_2 }_{1}}{2}\,, \qquad
c^{\phi_1 \phi^\dagger_2 \to \psi_1 \bar\psi_2}_{2}=c^{\phi_1 \phi_2 }_{1}\,.
\end{aligned}
\label{eq: cross identity}
\end{equation}
Thus, $c^{\phi_1 \phi_2 }_{1}$ and $c^{\phi_1 \phi_2 }_{2}$ may be chosen as independent coefficients, while the remaining coefficients are fixed by the Ward identity. The positivity requirement further implies that 
$c^{\phi_1 \phi_2 }_{2} - c^{\phi_1 \phi_2 }_{1} > 0$,
which in turn leads to the constraint
$c^{\phi_1 \phi^\dagger_2 \to \psi_1 \bar\psi_2}_{1} > 0$.  
In the non-supersymmetric case, however, this coefficient is unconstrained by positivity bounds. The relation Eq.~\eqref{eq: cross identity} therefore provides an additional restriction that arises solely from supersymmetry. This example consolidates the statement we claim in the previous section: an unconstrained positivity bound of a single component amplitude does not determine the positivity bound of the superamplitude. Instead, it gets promoted to a non-trivial one due to supersymmtry.

\subsection{$\mathcal N=1$ super Yang-Mills theory}
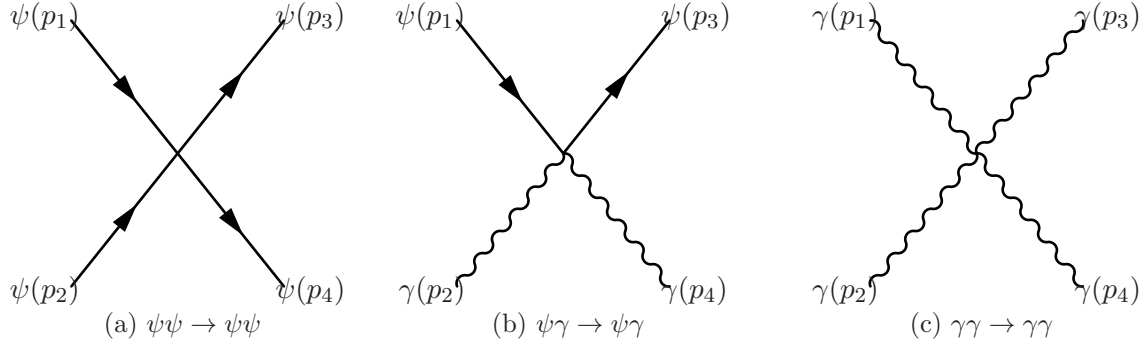
\begin{figure}[!t]
\centering

\begin{subfigure}{0.32\textwidth}
\centering
\begin{fmffile}{feyn10}
\fmfframe(2,8)(2,2){%
\begin{fmfgraph*}(100,100)
    \fmfleft{i1,i2}
    \fmfright{o1,o2}
    \fmf{fermion}{i1,w1}
    \fmf{fermion}{i2,w1}
    \fmf{fermion}{w1,o1}
    \fmf{fermion}{w1,o2}
    \fmfv{lab=$\psi(p_1)$,lab.dist=-0.05w}{i2}
    \fmfv{lab=$\psi(p_2)$,lab.dist=-0.05w}{i1}
    \fmfv{lab=$\psi(p_3)$,lab.dist=-0.05w}{o2}
    \fmfv{lab=$\psi(p_4)$,lab.dist=-0.05w}{o1}
\end{fmfgraph*}%
}
\end{fmffile}
\caption{$\psi\psi\rightarrow\psi\psi$}
\label{fig:contact-diagrams2a}
\end{subfigure}
\begin{subfigure}{0.32\textwidth}
\centering
\begin{fmffile}{feyn7}
\fmfframe(2,8)(2,2){%
\begin{fmfgraph*}(100,100)
    \fmfleft{i1,i2}
    \fmfright{o1,o2}
    \fmf{photon}{i1,w1}
    \fmf{fermion}{i2,w1}
    \fmf{photon}{w1,o1}
    \fmf{fermion}{w1,o2}
    \fmfv{lab=$\psi(p_1)$,lab.dist=-0.05w}{i2}
    \fmfv{lab=$\gamma(p_2)$,lab.dist=-0.05w}{i1}
    \fmfv{lab=$\psi(p_3)$,lab.dist=-0.05w}{o2}
    \fmfv{lab=$\gamma(p_4)$,lab.dist=-0.05w}{o1}
\end{fmfgraph*}%
}
\end{fmffile}
\caption{$\psi\gamma\rightarrow\psi\gamma$}
\label{fig:contact-diagrams2b}
\end{subfigure}
\hfill
\begin{subfigure}{0.32\textwidth}
\centering
\begin{fmffile}{feyn8}
\fmfframe(2,8)(2,2){%
\begin{fmfgraph*}(100,100)
    \fmfleft{i1,i2}
    \fmfright{o1,o2}
    \fmf{photon}{i1,w1}
    \fmf{photon}{i2,w1}
    \fmf{photon}{w1,o1}
    \fmf{photon}{w1,o2}
    \fmfv{lab=$\gamma(p_1)$,lab.dist=-0.05w}{i2}
    \fmfv{lab=$\gamma(p_2)$,lab.dist=-0.05w}{i1}
    \fmfv{lab=$\gamma(p_3)$,lab.dist=-0.05w}{o2}
    \fmfv{lab=$\gamma(p_4)$,lab.dist=-0.05w}{o1}
\end{fmfgraph*}%
}
\end{fmffile}
\caption{$\gamma\gamma\rightarrow\gamma\gamma$}
\label{fig:contact-diagrams2c}
\end{subfigure}

\caption{Four-point contact interaction diagrams for the component scattering processes involving the vector multiplet.}
\label{fig:contact-diagrams2}
\end{figure}

We now turn to the Abelian gauge theory, comprising the photon and its fermionic superpartner, the photino. Guided by the Ward identity, we examine the following contact interactions at mass-dimension eight: 
$\psi \psi \to \psi \psi$, 
$\psi \gamma \to \psi \gamma$, 
and $\gamma \gamma \to \gamma \gamma$. 
Following the same procedure as in the previous subsections, we note that the result for $\psi \psi \to \psi \psi$ can be directly adopted from earlier discussion. 
\begin{table}[!t]
\centering
\small

\begin{tblr}{
width=\textwidth,
colspec={|Q[c,m,0.11]|Q[c,m,0.46]|Q[c,m,0.46]|},
hlines,
vlines,
row{1}={font=\bfseries},
}

&
Operators
&
Amplitudes
\\

$\psi\psi\rightarrow\psi\psi$

&

{$
\mathcal{O}^{\psi \psi } 
= 
\big(\bar\psi \gamma^\mu\overset{\leftrightarrow}{\partial^\nu}\psi\big)
\big(\bar\psi \gamma_\mu\overset{\leftrightarrow}{\partial_\nu}\psi\big)\,.
$}

&

{$
\begin{aligned}
\mathcal{A}_{\mathcal{O}^{\psi \psi }}(\bar\psi,\bar\psi,\psi,\psi)
&=
-12 \frac{c^{\psi \psi }}{\Lambda^4}
\\
&\times s [1 2]\langle 3 4\rangle\,.
\end{aligned}
$}

\\

$\psi\gamma\rightarrow\psi\gamma$

&

{$
\mathcal O^{\psi\gamma}
=
i
\bigl(
\bar\psi\gamma^\mu\overset{\leftrightarrow}{\partial^\nu}\psi\bigr)
\bigl(B_{\mu\rho}B_\nu^\rho\bigr)\,.
$}

&

{$
\begin{aligned}
\mathcal A_{\mathcal O^{\psi\gamma}}
(\bar\psi,\gamma^+,\psi,\gamma^-)
&=
-\frac{2c^{\psi\gamma}}{\Lambda^4}
\\
&\times\langle14\rangle\langle34\rangle[12]^2\,.
\end{aligned}
$}

\\

$\gamma\gamma\rightarrow\gamma\gamma$

&

{$
\begin{aligned}
\mathcal O^{\gamma\gamma}_1
&=
(B_{\mu\nu}B^{\mu\nu})
(B_{\rho\sigma}B^{\rho\sigma})\,,
\\[0.6em]
\mathcal O^{\gamma\gamma}_2
&=
(B_{\mu\nu}B^{\mu\nu})
(B_{\rho\sigma}\tilde B^{\rho\sigma})\,,
\\[0.6em]
\mathcal O^{\gamma\gamma}_3
&=
(B_{\mu\nu}\tilde B^{\mu\nu})
(B_{\rho\sigma}\tilde B^{\rho\sigma})\,.
\end{aligned}
$}

&

{$
\begin{aligned}
\mathcal A(\gamma^+,\gamma^+,\gamma^-,\gamma^-)
&=
\frac{8c^{\gamma\gamma}_1}{\Lambda^4}\langle34\rangle^2[12]^2 \\
&+\frac{8c^{\gamma\gamma}_3}{\Lambda^4}\langle34\rangle^2[12]^2\,.
\end{aligned}
$}

\\

\end{tblr}

\caption{Dim-8 operators and the corresponding four-point amplitudes for the $\psi\psi\rightarrow\psi\psi$,
$\psi\gamma\rightarrow\psi\gamma$
and
$\gamma\gamma\rightarrow\gamma\gamma$
scattering processes (where $\tilde{B}^{\mu\nu} = \frac{1}{2}\epsilon^{\mu\nu\rho\sigma} B_{\rho\sigma}$ denotes the dual field strength tensor).}
\label{tab:vectoroperators}

\end{table}
Consider the three scattering processes shown in
Fig.~\ref{fig:contact-diagrams2}\subref{fig:contact-diagrams2a},
Fig.~\ref{fig:contact-diagrams2}\subref{fig:contact-diagrams2b} and Fig.~\ref{fig:contact-diagrams2}\subref{fig:contact-diagrams2c}, the corresponding independent dimension-eight operators and their associated four-point amplitudes are summarized below in Table~\ref{tab:vectoroperators}. The corresponding forward-limit positivity bounds therefore read 
\begin{equation}
c^{\psi\psi} > 0,\quad -c^{\psi\gamma} > 0,\quad c^{\gamma \gamma }_1 >0,\quad c^{\gamma \gamma }_3 > 0,\quad(c_2^{\gamma \gamma })^2 < 4c^{\gamma \gamma }_1c^{\gamma \gamma }_3.
\label{posgamma}
\end{equation}
The relations among the Wilson coefficients, which follow from the supersymmetric Ward identities, are given by
\begin{equation}
\mathcal{A}
(\bar\psi,\bar\psi,\psi,\psi) = -\frac{\langle 1 2\rangle}{\langle 1 4\rangle}\mathcal{A}
(\bar\psi,\gamma^+,\psi,\gamma^-)\,,
\end{equation}
\begin{equation}
\mathcal{A}
(\bar\psi,\gamma^+,\psi,\gamma^-) = -\frac{\langle 1 4\rangle}{\langle 4 3\rangle}\mathcal{A}(\gamma^+,\gamma^+,\gamma^-,\gamma^-)\,. 
\end{equation}
Thus we obtain 
\begin{equation}
3c^{\psi\psi} = - \frac{1}{2}c^{\psi \gamma } = 2 (c^{\gamma \gamma }_1 + c^{\gamma \gamma }_3)\,,
\label{eq: relative ratio vector}
\end{equation}
which is manifestly consistent with the positivity bounds Eq.~\eqref{posgamma}.

These relations can also be verified directly from the component expansion of the supersymmetric Lagrangian with four vector superfields $W,\overline W$,
\begin{equation}
    \mathcal{L}\supset \frac{c}{\Lambda^4}\int d^4\theta\, W^2\overline{W}^2= \frac{c}{3\Lambda^4}\mathcal{O}^{\psi\psi}-\frac{2c}{\Lambda^4}\mathcal{O}^{\psi \gamma }+\frac{c}{4\Lambda^4}\mathcal{O}^{\gamma \gamma }_1+\frac{c}{4\Lambda^4}\mathcal{O}^{\gamma \gamma }_3\,.
\end{equation}
As expected, the relative ratios of corresponding couplings agree precisely with Eq.~\eqref{eq: relative ratio vector} derived from the Ward identities.

\begin{table}[!t]
\centering
\begin{tabular}{|c|c|c|}
\hline
Number of independent parameters & Without supersymmetry & With supersymmetry \\
\hline
$\mathcal{A}(\bar\psi,\bar\psi,\psi,\psi)$ & 1 & 1 \\
\hline
$\mathcal{A}(\bar\psi,\gamma^+,\psi,\gamma^-)$ & 1 & 1 \\
\hline
$\mathcal{A}(\gamma,\gamma,\gamma,\gamma)(\text{all helicity sectors})$ & 3 &  1\\
\hline
$\text{all}$ & 5 &  1\\
\hline
\end{tabular}
\caption{Number of independent dim-8 operators with or without supersymmetry for the $\mathcal N=1$ super Yang-Mills theory.}
\label{tab:susy_operators3}
\end{table}

We have now addressed four of the five constraints in Eq.~\eqref{posgamma}. The last one, which involves the parity-odd operator $
\bigl(B_{\mu\nu} B^{\mu\nu}\bigr) \bigl(B_{\rho\sigma} \tilde{B}^{\rho\sigma}\bigr)$, only contributes to the amplitudes $(++++)$ and $(----)$. In the non-supersymmetric case, the coefficient $c^{\gamma \gamma }_2$ may take a nonzero value and will enter the resulting positivity bounds. Supersymmetry, however, imposes helicity selection rules analogous to the familiar MHV structure of gauge-theory amplitudes. In particular, the supersymmetric Ward identities forbid all-plus or all-minus four-point amplitudes for the vector multiplet, i.e., the amplitudes $(++++)$ and $(----)$ must vanish in a purely supersymmetric theory. 
\begin{equation}
\begin{aligned}
\mathcal A(\gamma^+,\gamma^+,\gamma^+,\gamma^+)
&=
\frac{8}{\Lambda^4}
\left(
c^{\gamma\gamma}_1-c^{\gamma\gamma}_3
-i c^{\gamma\gamma}_2
\right)\left([12]^2[34]^2+[13]^2[24]^2+[14]^2[23]^2\right)\,,
\\
\mathcal A(\gamma^-,\gamma^-,\gamma^-,\gamma^-)
&=
\frac{8}{\Lambda^4}
\left(
c^{\gamma\gamma}_1-c^{\gamma\gamma}_3
+i c^{\gamma\gamma}_2
\right)\left(\langle12\rangle^2\langle34\rangle^2
+\langle13\rangle^2\langle24\rangle^2
+\langle14\rangle^2\langle23\rangle^2\right)\,.
\end{aligned}
\end{equation}
For real Wilson coefficients, this implies,
\begin{equation}
c^{\gamma\gamma}_1=c^{\gamma\gamma}_3\,,
\qquad c^{\gamma \gamma }_2 = 0\,.
\end{equation} 

The numbers of independent Wilson coefficients contributing to the scattering amplitudes for the $\mathcal N=1$ super Yang-Mills theory are summarized in Table~\ref{tab:susy_operators3}.

\section{Extensions and future directions}\label{sec: future}
In this paper, we have investigated the positivity bounds in $\mathcal N=1$ supersymmetric theories. Starting from the most general massless four-point scattering processes and their on-shell amplitudes, we explicitly derive positivity constraints using supersymmetric Ward identities. By combining little-group scaling with crossing symmetry, we show that couplings associated with different scattering amplitudes are intrinsically related once supersymmetry is imposed. These relations are further demonstrated with explicit tree-level examples. While massive states can in principle be incorporated into this framework, taking the forward limit in Eq.~\eqref{eq:forward_limit_superamplitude} for massive particles is highly non-trivial and does not admit the same factorization as in the massless case.

We further note that four-point interactions are exceptional: imposing the supersymmetric Ward identities uniquely fixes the structure of each component amplitude, because the number of independent constraints exactly matches the number of unknowns. For higher-point processes, this balance no longer holds, as there are insufficient equations to fully determine all sub-amplitudes. Nevertheless, higher-point positivity has been explored in non-supersymmetric theories \cite{Chandrasekaran:2018qmx}, and whether the pattern proposed in this paper persists depends sensitively on the construction of the operator basis. This question is beyond the scope of the present work and will be addressed in future research.

Finally, extending this framework to loop levels is a natural next step, as the Ward identities remain valid to all orders by virtue of their algebraic nature. In non-supersymmetric theories, loop effects introduce additional diagrams and lower-dimensional interactions that must be accounted for. By contrast, one of the features in $\mathcal N=1$ supersymmetry is that fermion and boson loop contributions cancel, which is precisely what resolves the quadratic sensitivity responsible for the hierarchy problem. We therefore expect analogous cancellations to simplify the loop-level positivity analysis. The distinct behavior of loop-level positivity in supersymmetric versus non-supersymmetric theories is expected to be manifest in this setting, and we leave a detailed investigation of such cases for future work.

\appendix

\section*{Acknowledgements}
This work is supported by the National Natural Science Foundation of China (NSFC) under Grant No.\,12375091 and 12547137, and the Innovation Program for Quantum Science and Technology under grant No.\,2024ZD0300101. RW is also supported by the China Postdoctoral Science Foundation under Grant Number 2025M783371.

\section{Spinor-helicity formalism}
\label{app:spinor-helicity}

In this appendix we summarize the conventions and notation for the spinor-helicity formalism used in this paper. Our presentation follows the standard review by Elvang and Huang  \cite{Elvang:2013cua}.

The general solution to the Dirac equation takes the form
\begin{equation}
    \Psi(x) \sim u(p) e^{i p\cdot x} + v(p) e^{-i p\cdot x}\,,
\end{equation}
where $u(p)$ and $v(p)$ denote the positive- and negative-frequency modes, satisfying
\begin{equation}
(\slashed{p}+m) u(p)=0\,,\qquad(-\slashed{p}+m) v(p)=0\,.
\label{eq:Dirac-momentum}
\end{equation}
In the Weyl representation, we write the gamma-matrix as
\begin{equation}
\slashed{p} =
\begin{pmatrix}
0 & p_{a\dot b} \\
p^{\dot a b} & 0
\end{pmatrix}\,,
\end{equation}
where the bispinor components are defined as
\begin{equation}
p_{a\dot b} \equiv p_\mu (\sigma^\mu)_{a\dot b}
= 
\begin{pmatrix}
- p^0 + p^3 &  p^1 - i p^2 \\[4pt]
p^1 + i p^2 &  -p^0 - p^3
\end{pmatrix}\,,
\end{equation}
and similarly $p^{\dot a b} \equiv p_\mu (\bar{\sigma}^\mu)^{\dot a b}$. Here $\sigma^\mu = (1, \sigma^i)$ and $\bar{\sigma}^\mu = (1, -\sigma^i)$ with $\sigma^i$ $(i=1,2,3)$ the Pauli matrices.

For a massless momentum $p_\mu$ the on-shell condition is
\begin{equation}
    \det p= - p^{\mu}p_{\mu} = 0\,,
\end{equation}
so that $p_{a\dot{b}}$ is a rank-one $2\times2$ matrix and can be written as an outer product of two Weyl spinors:
\begin{equation}
    p_{a\dot{b}} = -\lambda_a \tilde{\lambda}_{\dot{b}}\,.
\end{equation}
We adopt the standard bra–ket notation
\begin{equation}
    \lambda_a \equiv |p]_a\,, 
    \qquad
    \tilde{\lambda}_{\dot{b}} \equiv \langle p|_{\dot{b}}\,.
\end{equation}
These bra–kets are nothing but two-component commuting Weyl spinors solving the massless Weyl equation. Indices are raised and lowered with the antisymmetric tensors $\epsilon^{ab}$ and $\epsilon^{\dot{a}\dot{b}}$:
\begin{equation}
    |p\rangle^{\dot{a}} = \epsilon^{\dot{a}\dot{b}}\langle p |_{\dot{b}}\,, \qquad [p|^{a} = \epsilon^{ab}|p]_{b}\,.
\end{equation}

For two lightlike vectors $p^\mu$ and $q^\mu$ we define the angle and square spinor brackets
\begin{equation}
    \langle pq\rangle \equiv  \langle p|_{\dot{a}}|q\rangle^{\dot{a}}\,, \qquad [pq] \equiv 
    [p|^a|q]_a\,.
\end{equation}
These brackets are antisymmetric,
\begin{equation}
    \langle pq\rangle = -\langle qp\rangle\,, \qquad [pq] = -[qp]\,,
\end{equation}
and mixed objects such as $\langle p|q]$ vanish in the massless construction. The scalar product of two massless momenta is given by
\begin{equation}
    2 p\cdot q = \langle pq\rangle [pq] = (p+q)^2\,.
\end{equation}
Momentum conservation for an $n$-particle amplitude (with all momenta taken outgoing), $\sum_{i=1}^{n}p^\mu_i=0$, implies the identity
\begin{equation}
    \sum_{i=1}^{n}\langle qi \rangle [ik]=0\,.
    \label{eq:momentum_conservation_spinor}
\end{equation}

As a concrete example, consider four-point scattering of massless external momenta $p_i$ $(i=1,\dots,4)$. The Mandelstam variables are defined as
\begin{equation}
    s = (p_1+p_2)^2\,,\qquad
    t = (p_1+p_3)^2\,,\qquad
    u = (p_1+p_4)^2\,,
\end{equation}
and, using $p_i^2=0$ and the spinor representation, they can be written as
\begin{equation}
    s = \langle 12\rangle[12]\,,  \qquad   t = \langle 13\rangle[13]\,,   \qquad   u = \langle 14\rangle[14]\,,
\end{equation}
with the usual constraint
\begin{equation}
    s+t+u=0\,.
\end{equation}
These relations provide the bridge between kinematic invariants and spinor-helicity variables, allowing one to trade Mandelstam variables for spinor products. This correspondence simplifies the analytic structure of amplitudes and underpins the compact supersymmetric representations used in the main text.

\section{Supersymmetric Ward Identities and Their Consequences}
\label{appendix: supersymmetric Ward Identity}
We now examine in detail the relations among amplitudes implied by the supersymmetry Ward identities. For definiteness, we focus on chiral superfields, since the structure of the Ward identities is identical for vector superfields and the extension to that case is straightforward. The action of the supersymmetry generators  on the annihilation operators is
\begin{align}
[Q , a^{-}(p)] &= |p]  b^{-}(p)\,, 
& [Q^{\dagger}, b^{-}(p)] &= |p\rangle  a^{-}(p)\,, \\
[Q , b^{-}(p)] &= 0\,, 
& [Q^{\dagger}, a^{-}(p)] &= 0\,, \\
[Q , b^{+}(p)] &= |p]  a^{+}(p)\,, 
& [Q^{\dagger}, a^{+}(p)] &= |p\rangle  b^{+}(p)\,, \\
[Q , a^{+}(p)] &= 0\,, 
& [Q^{\dagger}, b^{+}(p)] &= 0\,.
\end{align}

Given the manifest symmetry between the actions of $Q$ and $Q^{\dagger}$ is manifest, we may, without loss of generality, focus solely on $Q^{\dagger}$; the contribution from $Q$ produces symmetric terms and does not affect the final result. The Ward identity then gives
\begin{equation}
0 = \langle 0 | \big[  Q^{\dagger},  O_{1}(p_{1}) \cdots O_{4}(p_{4})  \big] |0\rangle\,.
\end{equation}
From this expression, we observe that, in order to obtain a non-vanishing relation among amplitudes, the number of $b^{-}(p)$ or $a^{+}(p)$ operators contained in the four $O_i$ must be exactly three.  

Cases with a single insertion are trivially excluded. For example, if only one operator $O_j$ fails to commute with $Q^\dagger$, the Ward identity reduces to
$\langle 0| O_1\cdots[Q^\dagger,O_j]\cdots O_4|0\rangle=0$, 
which contains only one term and vanishes identically. Two insertions are also excluded. In this case the commutator generates two terms, schematically $\langle 0|[Q^\dagger,O_i] O_j\cdots|0\rangle + \langle 0|O_i [Q^\dagger,O_j]\cdots|0\rangle=0$. By left-multiplying with a suitable spinor $\langle i|$, one may always choose kinematics such that only one of the two terms survives, yielding again one term which vanishes identically.

Configurations with four insertions similarly fail to yield nontrivial identities. This follows from the fact that the supercharge $Q^\dagger$ carries spin $1/2$. As a result, when the operator product $O_1(p_1)\cdots O_4(p_4)$ consists of four fermionic, four bosonic, or two fermionic and two bosonic operators, the action of $Q^\dagger$ necessarily produces amplitudes that vanish identically. This is a direct consequence of Lorentz invariance: any resulting scattering amplitude must be a Lorentz scalar and therefore cannot carry uncontracted Weyl spinor indices. Consequently, amplitudes involving an odd number of Weyl fermions must vanish.

The remaining possibilities can be illustrated by a representative example,
\begin{equation}
    0 = \langle 0 | \big[  Q^{\dagger},  b^{-}(p_1)b^{-}(p_2)b^{-}(p_3)a^{+}(p_4)  \big] |0\rangle\,.
\end{equation}
In this case, the commutator generates amplitudes containing three operators of the form $x^{-}$ and one operator of the form $x^{+}$ (or, equivalently, one $x^{-}$ and three $x^{+}$). Here $x$ may denote either the scalar operator $a$ or the fermionic operator $b$. Consequently, without loss of generality, one may impose an additional global $U(1)$ symmetry to exclude these amplitudes altogether.

Disregarding permutations of the external legs $1,2,3,4$, the only surviving relations are those involving the previously discussed amplitudes that respect crossing symmetry (the ones relevant for positivity), along with a single additional class of the form $\mathcal{A}_4(\phi^\dagger,\phi,\psi,\bar\psi)$. 
A representative relation among these amplitudes can be written as
\begin{equation}
\mathcal{A}
(\phi^\dagger,\phi,\psi,\bar\psi) = \frac{\langle 1 3\rangle}{\langle 1 4\rangle}\mathcal{A}
(\phi^\dagger,\phi,\phi,\phi^\dagger)\,,
\end{equation}
At the dim-8 level, both amplitudes are subject to positivity bounds without supersymmetry. The supersymmetric Ward identity is consistent with the bounds, but provides no additional nontrivial constraints. This situation changes when we consider two distinct chiral superfields, where we obtain the following relation:
\begin{equation}
\mathcal{A}
(\phi^\dagger_1,\phi_2,\psi_1,\bar\psi_2) = \frac{\langle 1 3\rangle}{\langle 1 4\rangle}\mathcal{A}
(\phi^\dagger_1,\phi_2,\phi_1,\phi^\dagger_2)\,,
\end{equation}
In this scenario, the Ward identity relates an inelastic scalar–fermion amplitude to an elastic scalar amplitude. Positivity of the latter therefore constrains the corresponding supersymmetric coefficient.

\bibliographystyle{utphys}
\bibliography{ref}

\end{document}